\documentstyle[12pt,aaspp4,epsfig]{article}

\begin{document}

\title{Optical Flash of GRB 990123: constraints on the physical parameter of the reverse shock}

\author{Y. Z. Fan,  Z. G. Dai, Y. F. Huang  and T. Lu}
\affil{Department of Astronomy, Nanjing University, Nanjing
210093, P. R. China } \affil{daizigao@public1.ptt.js.cn }

\begin{abstract}
The optical flash accompanying GRB 990123 is believed to be
powered by the reverse shock of a thin shell. With the best fitted
physical parameters for GRB 990123 (Panaitescu $\&$ Kumar 2001)
and the assumption that the parameters in the optical flash are
the same as those in the afterglow, we show that: 1) the shell is
thick but not thin, and we have provided the light curve for the
thick shell case which coincides with the observation; 2) the
theoretical peak flux of the optical flash accounts for only
$3\times10^{-4}$ of the observed.  In order to compensate this
divergency, the physical parameters the electron energy ration and
the magnetic ratio $\epsilon_e$, $\epsilon_B$ should be 0.61, 0.39
respectively, which are much different from those in the late
afterglow.
 \end{abstract}


\section{Introduction}
BeppoSAX ushered in 1999 with the discovery of a super-bright
$\gamma$-ray burst, GRB 990123. This GRB was intensively studied
by many groups world wide. At that time this burst was notable for
the richness of new results: the discovery of prompt optical
emission by ROTSE (Akerlof et al. 1999), the discovery of the
brightest optical afterglow  and its redshift $z=1.6004$ leads to
a huge energy release of $1.6\times10^{54}$ergs in $\gamma-$rays
alone (Briggs et al. 1999; Kulkarni et al. 1999a), and a break in
the optical afterglow light curve (Fruchter et al. 1999;
Castro-Tirado et al. 1999), and the radio flare (Kulkarni et al.
1999b). In the past three years, all of these phenomena have been
discussed in great detail. For instance, the steepening of the
$r$-band light curve from about $t^{-1.1}$ to $t^{-1.8}$ after two
days might be due to a jet which has transited from a
spherical-like phase to sideways expansion phase (Rhoads 1999;
Sari, Piran $\&$ Halphern 1999; Huang et al 2000 a, b, c, d; Wei
$\&$ Lu 2000). The steeping might also be due to a dense medium
which has slowed down the shock quickly to a non-relativistic one
(Dai $\&$ Lu 1999).

The most natural explanation for the strong optical emission
accompanying GRB 990123 is the synchrotron emission from a reverse
shock propagating into the fireball ejecta after it interacts with
the surrounding gas (Sari $\&$ Piran 1999; M\'{e}sz\'{a}ros $\&$
Rees 1999). Under this framework, the light curve of GRB optical
flash in a homogenous medium or in a stellar wind and its
corresponding Synchrotron self-Compton emission have been
discussed in great detail (Kobayashi 2001; Wang, Dai $\&$ Lu 2001
a, b; Wu et al. 2002; Fan et al. 2002). Several authors attempted
to constrain the intrinsic parameters, such as the Lorentz factor
of the shocked fireball ejecta relative to the unshocked fireball
ejecta ($\Gamma_{rs}$)(Wang, Dai $\&$ Lu 2000; Sari $\&$ Piran
1999). It should be noted, however, that these estimates were made
before accurate burst parameters for GRB 990123 were known, and
consequently they include approximations and parameters from other
GRB afterglows. Recently, by fitting the multi-frequency afterglow
light curves, physical parameters for eight GRBs, including GRB
990123 have been reported (Panaitescu $\&$ Kumar 2001, hereafter
PK01). This fitting has provided us the possibility to study this
unique event more quantitatively. With these parameters Soderberg
$\&$ Ramirez-Ruiz (2002, hereafter SR02) have estimated the
prompt reverse shock emission to be expected for these eight
bursts.

After a careful calculation with the parameters for GRB 990123 (we
assume these parameters in the optical flash are the same  as
those in the late afterglow), in section 2 we find the shell is
thick and we provide the adjusted light curve for the thick shell
case which coincides with the observation. In section 3, we find
the theoretical peak flux of the optical flash accounts for only
$3\times10^{-4}$ of the observed, if it is the reverse shock which
accounts for the optical flash. In order to compensate this great
divergency, the physical parameters of the electron energy ratio
and magnetic field energy ratio $\epsilon_e$, $\epsilon_B$ in the
optical flash phase should be much different from those in the
late afterglow phase.  In the final section we make some
discussions and give our conclusions.

\section{Light curves of the reverse shock emission for the thick shell case}
By fitting multi-frequency afterglow light curves, physical
parameters for 8 GRBs have been reported in PK01. Best fitted
parameters for GRB 990123 are: initial jet energy in afterglow
phase $E_{j,50}=1.5^{+3.3}_{-0.4}$ , initial opening angle
$\theta_0=2.1^{+0.1}_{-0.9}\rm deg$, environment number density
$n_{0,-3}=1.9^{+0.5}_{-1.5}$, $\epsilon_{e,-2}=13^{+1}_{-4}$,
$\epsilon_{B,-4}=7.4^{+23}_{-5.9}$, and the electron distribution
power-law index $p=2.28^{+0.05}_{-0.03}$.

The thin shell deceleration time, $t_{\gamma}$ can be estimated by
$t_{\gamma}\simeq 3E^{1/3}_{52}n^{-1/3}_{0,5}\eta^{-8/3}_{300}$
(Kobayashi 2000), where the parameters are scaled as
$E_{52}=E/10^{52}$, $n_{0,5}=n_0/5cm^{-3}$, $\eta_{300}=\eta/300$,
$\eta$ is the initial Lorentz factor of the fireball at the end of
the Gamma-ray burst, here we take its best estimated value
$\eta=900$ (SR02), $E$ is the isotropic energy of the fireball in
afterglow. With these parameters mentioned above we have
$t_{\gamma}\simeq8 \rm s<\Delta/c\simeq 20 \rm s$, where
$\bigtriangleup$ is the shell width. Therefore the shell is thick
but not thin. In fact if the shell is thin, the reverse shock will
be in sub-relativistic. However it is generally suggested that
$\eta \simeq 900$ to $1200$ (Wang, Dai $\&$ Lu 2000; SR02), and at
the reverse shock crossing time $\Gamma$, the Lorentz factor of
the fireball $\simeq 300$ (PK01), i.e. $\Gamma_{rs}\simeq 5/3$ to
$2$ which is mid-relativistic, so the shell should be thick. This
conclusion coincides with the result of Wang, Dai $\&$ Lu (2000).
Some authors argued that if the shell was thick, the theoretic
light curve would be much different from what one observed (Sari
$\&$ Piran 1999; Kobayashi 2000; SR02). Below we investigate that
problem.

In the thick shell case, the reverse shock crosses the shell at
$T\simeq\Delta/c$. At the reverse shock crossing time $T$ the
break frequency $\nu_m$ and the peak flux are
\begin{equation}
\nu_m=10^{13} ({p-2\over p-1})^2({\epsilon_e\over0.1})^2
({\epsilon_B\over10^{-3}})^{1/2}({ n\over10^{-2}})^{1/2}
({\Gamma_A\over300})^2(\Gamma_{rs}-1)^2{1\over(1+z)}{\rm Hz},
\end{equation}
\begin{equation}
F_{\nu m}=1.2\times10^{-2}({D\over
10^{28}})^{-2}{N_e\over10^{52}}({\Gamma_A\over300})^2({n\over10^{-2}})^{1/2}
({\epsilon_B\over10^{-3}})^{1/2}(1+z) {\rm Jy},
\end{equation}
where the relations
$\gamma_m=(p-2)/(p-1)(m_p/m_e)\epsilon_e(\Gamma_{rs}-1)$ for
$p>2$, $F_{\nu m}={N_e \Gamma_A P_{\nu m}(1+z)\over 4\pi D^2}$,
$P_{\nu m}=\phi_p {\sqrt3 e^3 B \over m_e c^2}$ and $B=3.9\times
10^{-2} n_1^{1/2} ({\epsilon_B/10^{-2}})^{1/2} \Gamma_A $ have
been used, $\Gamma_A$ is the Lorentz factor of the shocked shell,
$\Gamma_{rs}$ is approximated by $(\Gamma_A/\eta+\eta/\Gamma_A)/2$
for $\Gamma_A$, $\eta\gg1$, $N_e$ is the total number of electrons
in the shell, $D$ is the luminosity distance (we assume $H_0=65
\rm km$ $\rm s^{-1}$ $\rm Mpc^{-1}$, $\Omega_M=0.3$,
$\Omega_\wedge=0.7$), $z$ is the redshift of the burst, $\phi_p$
is a function of $p$, whose value is $\sim 0.6$ for $p\sim 2.28$
(Wijers $\&$ Galama 1999).

The scalings before and after $T$ in the homogenous medium case
have been discussed by Kobayashi (2000). A difference between
Kobayashi's and our scalings is: at early times the reverse shock
is Newtonian (Kobayashi assumed it was relativistic), so
$\Gamma_{rs}-1\propto \Gamma_A^2 f^{-1}$, $\Gamma_A\simeq\eta$
(Sari $\&$ Piran 1995). In the thick shell case:  the spreading is
not important, then $f\equiv{n_4 \over n_1}\propto R^{-2}$. Noting
$R\sim2\Gamma_A^2tc$, we have $f\propto t^{-2}$, i.e.
$\Gamma_{rs}-1\propto t^2$. Substituting this relation into
equation (1) we obtain $\nu_m\propto t^4$. Noting $N_e(t)\propto
t$ (Kobayashi 2000) and substituting this relation into equation
(2), we have $F_{\nu m}\propto t$. For $\nu_m<\nu<\nu_c$ we have
$F_{\nu}\propto t^{2(p-1)}F_{\nu m}\propto t^{2p-1}$.

For $\Gamma_{rs}\gg 1$, $(\Gamma_{rs}-1)^2\Gamma_A^2\sim \eta^2/4
$, equation (1) reduces to $\nu_m\sim \rm constant$, as the
case suggested by Kobayashi (2000). Combining Kobayashi's results
and ours we get the flux at a given frequency $\nu$, for
$\nu_m<\nu<\nu_c$
\begin{equation}
 F_{\nu}(t<T) \propto \left\{
\begin{array}{lll}
   t^{2p-1}, & {\rm for}\,\, \Gamma_{rs}-1\ll1, \\\
   t^{1/2}, & {\rm for}\,\, \Gamma_{rs}-1\gg1,
   \end{array} \right.
\end{equation}
\begin{equation}
F_{\nu}(t>T)\propto t^{-(73p+21)/96}.
\end{equation}
The observed optical light curve of GRB 990123 at early times shows
a fast rise and a slower decay, whose power-law indices are
3.3 and -2.0 respectively. On the other hand, for $p=2.28$ we have
$2p-1=3.56$ and $-(73p+21)/96=-1.95$, then we expect that the
light curve rises faster at early times (for a power-law index 3.56)
then slowly (for a power-law index 0.5) before it reaches its peak.
Unfortunately, the lack of data for the optical flash prevents us
to check it more quantitatively. By now we have successfully
explained the fast rise of $t^{3.3}$ and slow decay of $t^{-2.0}$ in
the thick shell case.

\section{The expected peak flux of the optical
flash } With the best fit parameters of GRB 990123 afterglow for a
homogeneous medium with $90\%$ confidence level, we have (see in
PK01 ): $M_{jet}\simeq 0.28 \times 10^{-6}M_\odot$, $\Gamma_{0}
\simeq 300$. Correspondingly, $N_e$ and $\Gamma_A$ in equation (2)
are $N_e\simeq 5 \times 10^{53}$, $\Gamma_A \simeq 300$
respectively. The synchrotron spectrum for $\nu_m< \nu_{obs}<
\nu_c$ is given by
\begin{equation}
F_{obs}=F_{\nu m}(\nu_{obs}/\nu_m)^{-(p-1)\over2}.
\end{equation}

Substituting equations (1) and (2) into equation (5) we have
\begin{equation}
F_{obs,peak}=0.012\times[0.14({p-2\over p-1})({\epsilon_e\over
0.1})(\Gamma_{rs}-1)]^{p-1}[({\epsilon_B\over10^{-3}})({n\over10^{-2}})]^{p+1\over4}
({D\over10^{28}})^{-2}{N_e\over10^{52}}(1+z)^{{3-p\over2}}({\Gamma_A\over
300})^{p+1} {\rm Jy}.
\end{equation}
When $\nu_{obs}=5\times 10^{14}{\rm Hz}$,  $\Gamma_{rs}-1\simeq 1$
and other best fitted parameters of GRB 990123  have been taken
into calculation, we have $F_{obs,peak}=3\times 10^{-4}${\rm Jy},
which is much less than what we observed, $F_{peak} \simeq 1$Jy
(Akerlof et al. 1999).

One may argue that if the optical flash was born in a dense
envelope, for instance $n \simeq 40 \rm cm^{-3}$, the divergency
will disappear. However, there is no more evidence for that.
Another way is to assume that the physical parameters in the
optical flash are different from those in the late afterglow, for
example $\epsilon_e \simeq 0.61$, $\epsilon_B \simeq 0.39$ ($n$ is
the same as that in afterglow phase ) can compensate this
discrepancy safely. But this means in different phases (the GRB,
very early afterglow and the late afterglow) the physical
parameters may be much different.  In fact, as early as in 2000,
it has been proposed that the high energy spectral power-law
indices ($\beta$) for GRBs 970508, 990123, 990510, 991216 are
-1.88, -2.30, -2.49, -2.00 respectively (Fenimore $\&$
Ramirez-Ruiz 2000), i.e., corresponding $p$ in the GRB phase are
1.76, 2.60, 2.98, 2.00 respectively. However the best fitted $p$
in the afterglow phase are 2.18, 2.28, 1.83, 1.36 respectively for
these four GRBs (PK01). Obviously they are quite different.

Dai $\&$ Lu (1999) have proposed the dense medium model to explain
the afterglow decay of GRB 990123. The parameters derived from
that model are $\epsilon_e\sim 0.1$, $\epsilon_{B,-6}\sim0.02$,
$n\sim3\times10^6$. In this case, if we set $\Gamma_A\simeq 300$,
$N_e=E_{iso,\gamma}/\Gamma_0 m_pc^2$, $p=2.3$, we have
$F_{obs,peak}\simeq 1 \rm Jy$. However, according to the jump
conditions of the shock, the Lorentz factor of the shocked shell
should approximately equal to that of shocked ISM. The Lorentz
factor of the forward shocked ISM could be obtained from the
standard afterglow model (e.g. Sari, Piran $\&$ Narayan 1998):
$\Gamma_{A,fs}(t)\simeq6({E_{52}\over
n})^{1/8}({t_d\over(1+z)})^{-3/8}$.  For $E_{52}\sim 22$ and
$n\sim3\times10^6$, we have
$\Gamma_{A,rs}(50s)=\Gamma_{A,fs}(50s)\simeq 32$, which is much
below 300. From equation (6), such small $\Gamma_{A,rs}$ will lead
to a much smaller $F_{obs,peak}$ than the observed. This negative
result favors our opinion that these parameters for later forward
and early reverse shocks are different, at least in the case of
GRB 990123.

\section{Summary and discussion}
With the parameters for GRB 990123 provided in PK01, we have shown
that the shell is thick but not thin. The adjusted light curve for
the thick shell case can account for the observed light curve of
the optical flash of GRB 990123. However the expected peak
emission flux is much less than the observed. The parameters
derived from the dense medium model by Dai $\&$ Lu (1999) have
been considered, too, but the expected peak emission is still much
less than observation. If the optical flash was really produced by
the reverse shock, the parameters $\epsilon_B$, $\epsilon_e$, even
$p$ in optical flash should be much different from that in the
late afterglow. Unfortunately there is no enough data for us to
study it more quantitatively. New observations are needed to
provide us a chance to understand optical flashes in more detail.

With eight GRBs' parameters, SR02 have estimated the reverse shock
peak emission for seven bursts---for reasonable assumptions about
the velocity of the source expansion, a strong optical flash $m_V
\sim 9$ was expected from the reverse shock, then the best
observational prospects for detecting these prompt flashes are
high-lightened. It is easy to see that  equation (6) in this note
provides similar results. For instance: for GRB 000926, we have
$F_{obs,peak}\sim0.2(\Gamma_{rs}-1)^{p-1}\rm Jy$. Surprisingly,
although many researchers have tried their best, there is no more
optical flashes that have been observed (Akerlof et al. 2001;
Kehoe et al. 2001). SR02 suggested that the dust obscuration
seemed to be the most likely reason for non-detection. However,
considered the discrepancy between the observed peak flux and the
theoretically expected value, the reverse shock emission might be
insignificant, more reliable model to explain that "unique"
observation is needed.

\acknowledgments We would like to  thank D. M. Wei , X .Y. Wang,
X. F. Wu and Z. Li for valuable discussion. This work was
supported by the National Natural Science Foundation of China, the
National 973 project (NKBRSF G19990754), the Special Funds for
Major State Basic Research Projects, and the Foundation for the
Author of National Excellent Doctoral Dissertation of P. R. China
(Project No: 200125).

\end{document}